# Vox Populi, Vox ChatGPT

## Large Language Models, Education and Democracy

Niina Zuber[1] & Jan Gogoll[2]

Abstract

In the era of generative AI and specifically large language models (LLMs), exemplified by ChatGPT, the intersection of artificial intelligence and human reasoning has become a focal point of global attention. Unlike conventional search engines, LLMs go beyond mere information retrieval, entering into the realm of discourse culture. Its outputs mimic well-considered, independent opinions or statements of facts, presenting a pretense of wisdom. This paper explores the potential transformative impact of LLMs on democratic societies. It delves into the concerns regarding the difficulty in distinguishing ChatGPT-generated texts from human output. The discussion emphasizes the essence of authorship, rooted in the unique human capacity for reason - a quality indispensable for democratic discourse and successful collaboration within free societies. Highlighting the potential threats to democracy, this paper presents three arguments: the Substitution argument, the Authenticity argument, and the Facts argument. These arguments highlight the potential risks that are associated with an overreliance on LLMs. The central thesis posits that widespread deployment of LLMs may adversely affect the fabric of a democracy if not comprehended and addressed proactively and properly. In proposing a solution, we advocate for an emphasis on education as a means to mitigate risks. We suggest cultivating thinking skills in children, fostering coherent thought formulation, and distinguishing between machine-generated output and genuine, i.e. human, reasoning. The focus should be on responsible development and usage of LLMs, with the goal of augmenting human capacities in thinking, deliberating and decision-making rather than substituting them.

# Introduction

---

[1] Bavarian Institute of Digital Transformation, Gabelsbergerstr. 4, 80333 Munich
[2] Bavarian Institute of Digital Transformation, Gabelsbergerstr. 4, 80333 Munich

The latest innovation in artificial intelligence, large language models (LLM) such as ChatGPT, have attracted worldwide attention this past year. ChatGPT transcends conventional search engines, such as Google or Bing, by emulating the capacity for human reasoning. Whereas search engines assist in accessing and storing information - imitating our memory - generative AI aims at our discourse culture. It is no longer purely about access to factual knowledge. ChatGPT's outputs are information tokens processed into complex, often extensive texts that give the impression to be well-considered, independent opinions on a more or less demanding question. ChatGPT presents itself as a wise person, providing answers to almost every question you might have (Stahl & Eke, 2023). Thus, not only is the content represented as factual knowledge in the form of propositional sentences based on propositional logic, the text is also presented in the form of an argumentation and thus shows the characteristic traits of reasoning. In May 2023, the Future of Life Institute advocated for a six-month pause to comprehend the risks posed by such systems and to instigate any required regulations. Among the signatories are well-known technology advocates such as Elon Musk and Yuval Harari. This controversial moratorium is intended to delay the deployment of models more powerful than GPT 4 for at least six months allowing for introducing mechanisms that address societal and ethical concerns. Inventions such as ChatGPT promote the concern that AI might outstrip human capabilities and so nurtures a dystopian vision where intelligent machines could supplant, or at the very least marginalize, humans, and potentially assert dominance over them in the future (Bostrom, 1998).

As automation continues to encroach upon human domains, it prompts us to contemplate the implications of this transition. While the displacement of manual labor by machines is a longstanding phenomenon, contemporary computer technologies are progressively entering into domains intrinsically associated with human attributes, such as the ability to reason and our existence as an emotional and social living being (Agar, 2019). Acknowledging that these capabilities are fundamental for successful collaboration among humans is crucial. As such the capacity for respectful discourse is essential for the sustainability of a functioning democratic system. As we will discuss, LLMs have the potential to augment communication proficiencies, yet we need to ponder how they transform the very way we reason. Stahl and Eke (2023) highlight that "[...] the moral benefits of ChatGPT [...] remain somewhat fuzzy, there are more clearly defined concerns. The most prominent set of those has to do with the fact that ChatGPT texts can be difficult to distinguish from human output which can lead to problems of attribution of authorship." However, the concept of authorship entails more than the assignment of an output to a person which is widely discussed as a copyright issue. The essence of human nature lies in our unique capacity of being the origin of our statements and actions. This aptitude, which we refer to as reason (*Vernunftfähigkeit*), as we shall elucidate, necessitates substantial practice and is not readily replaceable, nor should we readily entertain the notion of such substitution. Above all, our social life relies on the essential skills of understanding ourselves as self-determinant beings. Thus, they are even vital for democracies: Without an understanding of others, we cannot conduct a well-founded exchange. The necessary attitudes required for a successful exchange among free and equal people cannot simply be delegated to machines, just as normative reasoning cannot (Nida-Rümelin, 2018). In order to effectively emphasize a required shift in skill development, it is imperative to gain a comprehensive understanding of the fundamental foundations of the concept of authorship. As Dwivendi et al. (2023) state: "The various domains that are affected by ChatGPT should be clearly articulated and the disruptions clearly understood. This will then lead to a better delineation of what the new world looks like and the skills needed to function in that new world."

In the subsequent discussion, we will elucidate three distinct arguments for why an excessive dependence on Large Language Models could potentially pose a threat to a well-functioning democratic society that relies on informed and empowered citizens. First and foremost, we will delve into the pivotal component of a thriving democracy, namely, individual autonomy, which is fundamentally rooted in the exercise of judgment. Secondly, it is imperative to underscore that a democratic society serves as the societal framework best suited for nurturing individual autonomy. Following this, we will explore three arguments—namely, the Substitution argument, the Authenticity argument, and the Facts argument—that serve as a framework for assessing the risks and benefits associated with the deployment of LLMs within democratic practices.

These arguments establish the foundation for our central thesis, positing that the wide-spread deployment of Large Language Models, despite potential advantages, might exert an adverse impact on the fabric of democracy. This may occur if the repercussions of excessive reliance on AI are not comprehended, and proactive measures are not implemented to address them.

Finally, we argue that in order to mitigate the potential risks associated with widespread Large Language Model use, a crucial step is to emphasize education. This involves training children to cultivate thinking skills rather than solely relying on LLMs, fostering the ability to formulate thoughts coherently to enhance judgment, and conveying the distinction between machine-generated output and genuine reasoning. By promoting responsible development and usage of LLMs, the focus should be on augmenting human capacities in deliberation rather than substituting them.

## Power of Judgment is a human capability and skill

Reason has been considered to be a distinctive attribute of human beings - the main differentiator that sets us apart from other lifeforms on earth. It seems to be a necessary element of the *Conditio humana*. No matter if we argue about facts (*theoria*) or actions (*praxis*). Software systems trying to imitate and simulate human thinking is nothing new, but with ChatGPT the shift is novel to the extent that generative AI aims at the capability of formulating own arguments based on facts and applied logic. Adopting ChatGPT's arguments may degenerate human reasoning into a mere expressing of opinions, as if reflecting and arguing would be all about fitting arguments into narratives. However, the act of reasoning involves the capability to assess arguments in favor and against a particular position, while reasons themselves are comprehended in an objective manner: reasons are not merely subjective states, like preferences or pro-attitudes, rather they speak in favor of judgements and decisions. Reasoning is the process of discerning which reasons objectively prevail, thereby determining what constitutes a rational belief or decision. We give reasons in order to convince and let oneself be convinced by the best available arguments  (Nida-Rümelin, 2023)**.**

This anthropological premise is at the core of Digital Humanism (Nida-Rümelin, 2018; Nida-Rümelin & Weidenfeld, 2022): The autonomous individual perceives and undergoes agency through the act of presenting and adhering to reasons. This person is not guided solely by states and circumstances that she desires to optimize but, instead, adheres to the superior objective reasons (Nida-Rümelin, 1991, 2023; Wedgwood, 2017). Furthermore, the autonomous person is capable of revealing those reasons and acting in accordance. On the basis of this ability, we attribute a talent for reason to persons. Together we practice the use of reason, evaluate the better reasons and come to agreements or compromises. We train this ability at schools and by participating in social structures which are designed to ensure and encourage our agency.

Kant's differentiation between understanding, judgment, and reason sheds light on why reason is considered a distinctive characteristic of human agency (Kant 1987/1790, Introduction). The faculty of understanding empowers us to derive concepts from sensory impressions and subsequently analyze the relationships between objects. Understanding is primarily descriptive and contingent, as its content is determined by perceptual entities and their interrelations. It is the mental function responsible for organizing the sensory data according to recognized rules, facilitating the formulation of terms, rules, and concepts. This cognitive endeavor distinguishes itself from the faculty of judgment, as it identifies and situates the individual case — whether a particular situation, action, or object — within a framework of established rules. The acquisition of the faculty of judgment is a formidable challenge, particularly within the context of democratic structures. The outsourcing of this critical ability presents a tangible threat to the fundamental principles of a free and egalitarian society. It can be assumed at first that arithmetic thinking or outsourcing knowledge to encyclopedias or to the spheres of the internet, can lead to similar fears of loss of an essentially human way of life. However, it should be emphasized that the power of judgment remains the skill needed to find one's way in a complex, constantly changing world: It is not limited to one area, it does not aim at ordering a certain form of interrelation, but it is rather constitutive to our daily life-form (Strawson, 1974; Mead, 1934). The power of judgment shapes our normative claims and gives us orientation.

Power of judgment discerns the appropriateness amidst the chaos of sensory experiences through the application of conceptualizations supplied by reason. Reasoning, in turn, necessitates the evaluation of values and the extraction of sound arguments from the interplay between interpersonal understanding and personal judgment. This dynamic operates in both theoretical judgments and practical actions. There exists no universal algorithm for determining the rationality of a thought or action. Consequently, figures like Wittgenstein and proponents of American pragmatism emphasize that language acquisition is an immersion in life. The practical and pragmatic aspects of reasoning emerge from the crucible of lived experiences, requiring validation within that realm.

We refer to our ability to react adequately to a new situation as *phronesis* (Aristoteles) or *power of judgment* (Kant). Commonly, this kind of practical thinking and acting is presented in the form of a practical syllogism comprising a normative statement, followed by a descriptive statement to conclude in yet another normative statement. Kant (1987/1790) distinguishes two different types of thinking in regard to practical reasoning. Firstly, he highlights the *bestimmende Urteilskraft (determinant judgment)*. In that case, we are aware of the normative principles that are relevant to the situation. However, we need to subsume the empirical special individual case in respect to the known normative principle. For example, we know that plagiarism is undesirable and in many cases even a legal offense, but we do not yet understand how the output of generative AI is related to that normative principle. Secondly, Kant then introduces the *reflektierende Urteilskraft (reflective judgment)* (Kant, 1987/1790, XXV, p.19)*.* This entails the practical thinking faced with a single individual case, while we lack the corresponding normative principle by which we are supposed to evaluate that case. In such circumstances we might argue that we know how to formulate different prompts for generative AI systems to gain an optimal output, but yet we do not know whether these prompts should be formulated according to some ethical standard and, if so, how they are to be applied correctly.

Sometimes this practical endeavor requires an intensive cognitive undertaking, other times we respond almost automatically. This depends on the situation, whether it is a daily routine or whether we are confronted with a novel situation and how easily comprehensible this new situation is. Most importantly though Kant realized that if the power of judgment applies logical rules then it cannot itself be understood in terms of logical rules. And if this power cannot be described in a rule-based terminology then it has to be grasped as a talent. However, talents cannot be acquired by the study of rules but need training. Thus, the ability of judging must be exercised: Only through practical experience of patterns of agency do we develop a sense of the rational structure of a situation. In similar situations, we can thus act appropriately, without explaining exactly what the normative claim is supposed to be.

While LLMs may seem to generate judicious and intelligent statements, they lack a human-like sense of understanding. Nonetheless, it is their outward semblance that gives rise to the impression of engaging with a rational entity, a notion that ultimately proves indispensable for elucidation. It is paramount to underscore the fundamental difference between human judgment and LLMs to understand how to use and handle such machines responsibly. As we demonstrated, human judgment is distinguished by its capacity for complex contextual comprehension. Human beings possess an ability to delve deeply into intricate situations perceiving various factors, such as cultural nuances, emotional connotations, and individual experiences. For example, a single action becomes reasonable by its integration into a series of actions, i.e. practice - whether this practice is a game, a profession or a social convention (Nida-Rümelin, 2019; 2023). This practice serves as a configuration of the single action (pragma). This is what Wittgenstein had in mind when he introduced the concept of Sprachspiel (Language-game) (Wittgenstein, 1953). Within Sprachspielen, we learn directly when an utterance or an action is inaccurate or inappropriate in that context, or even morally reprehensible. We know that the rules exist, but cannot make them explicit, explain them through a chain of other rules (Kripke, 1982) or readily apply the skills it requires, but must practice them. Skill arises from practice in a repeated intentional doing, i.e. training. From doing so habit arises, then skill and attitude (ethos) (Brinkmann 2021, p.24 ff.). Only then can we talk about a skill that we can apply correctly within a certain setting. This adaptability is a fundamental feature of human judgment, enabling individuals to respond effectively to changing circumstances. In stark contrast, LLMs calculate the probabilities of further word sequences depending on the command/prompt on an existing data basis. Participating in a life world, a world of feelings, what we call organic life, leads to an understanding of language beyond probable word contexts.

LLMs do not understand the person behind its utterance, it does not understand the emotional or mental states the person is in (Nida-Rümelin, 2018). This is yet the basis of successful communication. Successful communication depends on the adherence to the rules of truthfulness, trust and reliability (Nida-Rümelin, 2011). In our communication, we acknowledge and assume that our partner adheres to these standards. We also expect them to use their ability to reason, attributing this ability to them, and we honor their expressions as reflections of their individual selves. Communication holds significance not solely in the expression of beliefs but, more crucially, as a medium for aligning ourselves within the world. In other words, communication is intrinsically tied to reality. Hence, human judgment diverges significantly from LLMs as its role is in fostering intersubjective dialogue. Collaboratively figuring things out, discovering a path, and articulating possibilities for coexistence—these are tasks beyond the capabilities of a machine. Recognising a compromise and wanting to push through a different argument, understanding why the other person holds certain views and attitudes, that is something a machine cannot capture. A machine does not encompass the capacity to reflect oneself in its counterpart and attain awareness of that reflection. This practical wisdom is a distinctive characteristic of human judgment. It is an accumulation of life experiences, knowledge, and adaptive learning, creating a reservoir of wisdom that empowers individuals to navigate the complexities and uncertainties of real-world situations. This dimension is conspicuously absent in LLMs, which lack the depth of experiential wisdom that human judgment encompasses.

Yet, it is precisely this ability to debate, which is based on our ability to reason, that is essential for our social life in democratic societies as free and equal citizens. Only if we all meet as individuals gifted with these skills can we achieve a respectful exchange. This gives rise to certain normative educational requirements. Thus, we need to take our authority seriously, discerning how we can enhance it with the use of LLMs without compromising its integrity.

## Democracy as a life-form

There is an analytical link between the practice of giving and taking reasons and the functioning of democracy: The ability to argue for one's convictions, to let others demonstrate their reasons, and come to terms together are seen as the core of the democratic practice of public reasoning. According to pragmatist and educational theorist John Dewey, democracy is a procedure for reaching agreement on the good of society through the exchange of reasons (Dewey, 1903). As is the case with all skills, this human capability of giving and taking reasons necessitates practice: Guided practice integrated into education and upbringing.

Democracy can be grasped as collective self-determination of equal and free humans. There is no order of domination by nature, no domination of feudals over subjects, of men over women, of highly developed cultures over so-called primitive ones, there are no castes and estates that legitimately assign human individuals their role in society and the state. Assignments of this kind shaped human history and continue to do so today. They are illegitimate (Nida-Rümelin, 2020).

As citizens of a democracy, people are not factually equal; they differ by social and cultural origin, by gender, by abilities, interests and talents, by body size and appearance, but at the same time they are sufficiently similar not to accept any order of rule by nature. To put it differently: Reasonably, only the consent of the free and equal instigates a legitimate political order. It has two sources: the necessity of common, collective decisions, the indispensability of a legal order that secures individual rights and enables collective decision-making, and the common desire for collective self-determination.

So to speak, democracy is not merely a process of aggregating preferences. Democracy is based on consensus and shared normative convictions, at the center of which is the idea of autonomy and equality for all people. These norms extend beyond the political realm; they are not confined to regulations governing citizens' interactions but are fundamental to the entirety of the democratic lifeform. The economic and cultural practice is characterized by different norms. There is a close connection between individual autonomy and democratic collective autonomy. Under the assumption of individual autonomy, only those collective decisions and state institutions are permissible which are based on consensus. Consensus here does not mean the respective concrete agreement to a decision or a state institution, but a higher-order consensus that refers to the procedures that legitimize a political decision or the existence of an institution of the state. Our understanding of individual and democratic autonomy is nothing other than a particular reading of the freedom and equality of all human beings postulated by modern political theorists, starting with Thomas Hobbes (1651/1998). It is the idea that human beings are equally capable and entitled to live a life according to their own ideas, combined with the demand that every human individual be accorded the necessary equal respect so that this capacity can develop and the individual right to self-determination is not violated.

Nevertheless, our focus lies not in empirical inquiries but in normative considerations: we are not concerned with determining whether consensus turns out to be favorable for securing a group's ability to act, but rather what justifies collective decisions. If we assume that each group is made up of independently autonomous individuals, collective action is deemed acceptable only when rooted in consensus, even if that consensus is confined to decision-making procedures. We are then dealing with a consensus of a higher order. Without such a consensus, however, no one can be forced to participate in the procedures of collective decision-making in order to secure collective agency - such a consensus is thus constitutive of collective self-determination. Democracy is described as a specific form of collective autonomy in which individual autonomy is preserved.

Cooperation is the great challenge of democratic practice. In order to be able to cooperate, people must be able to transcend their own point of view of interests to such an extent that they can meet other points of view of interests with respect and include them in a cooperative practice. Cooperation does not require convergence of opinions and interests, but mutual recognition as equals and free, the willingness to exchange reasons and to coordinate across different points of interest to the extent that cooperation is possible. Rawls (1993) noted that "a plurality of reasonable, yet incompatible, comprehensive doctrines is the normal result of the exercise of human reason within the framework of the free institutions of a constitutional democratic regime." Briefly, Rawls, along with many others, contends that the institutions of modern democracies, rooted in toleration and recognition of what economists term bounded rationality, as well as what Rawls labeled the burdens of judgment, will inevitably give rise to a multitude of diverse beliefs and moral perspectives. Transcending one's own interests and opinions does not mean abandoning them, but taking them into account in joint practice with other interests and opinions. The criteria of appropriate reconciliation are themselves part of political practice; they cannot be established externally, by ethics, philosophy, social science, by expertise and experts. These merely contribute to conceptual and normative clarification. They can help to assess the expected consequences and side-effects, they can draw attention to value conflicts, direct attention to phenomena that are neglected in public discourse, give a voice to minorities but also to silent majorities. Yet, they cannot replace the political process itself for that would be the end of democracy. At its essence, democracy rests on a civic foundation, wherein the pivotal element is the capacity for autonomous and equal citizens to engage in discourse; without this, democracy undergoes erosion.

As we have tried to show, the talent of reason is a skill that requires practice. As the practice becomes more sophisticated, it may evolve into an ethos, serving as the foundation for a community, a collective life-form. This foundation is based on an independent use of reason and of common discourse. One such community that promotes and demands this skill is the democratic society. Nevertheless, the consequences of an excessive reliance on LLMs remain a subject of concern. Exploring the ways in which LLMs can be harnessed to uphold our identity as free and equal individuals is of paramount importance. Consequently, we should scrutinize the impact of ChatGPT on our comprehension of democracy and our autonomous selves. In our perspective, three categories can be established to classify the various use cases of extensive LLMs utilization as threats to democracy: Erosion of Skills, Erosion of Authorship, and Proliferation of Mis- and Disinformation. We will now review and discuss each of these three arguments in more detail.

Losing Skills via Substitution

In this section we will approach the losing skills via substitution argument regarding a potential decline in the ability to reason if we use LLMs for argumentation purposes and its effect on the democratic life-form. Considering the normative educational imperative to engage our reasoning abilities, crucial for us as autonomous individuals, it becomes imperative to deliberate more thoughtfully on how machines can be judiciously employed to assist us in our journey as self-determining human beings. The notion that technological advancements contribute to a decline in skill acquisition is far from novel. It dates back to the invention of the printing press and the introduction of calculators, freeing people from the need to engage in laborious calculations. This trend extends to the use of recipe books in culinary practices and the general dependence on search engines and the internet. The latter has diminished the imperative nature of retaining factual information, thereby potentially compromising our memory capabilities—a phenomenon some have referred to as the "Google effect" (Rowlands et al., 2008). There seems to be a worry among many, as Nyholm (2023) observes, that AI "might make us less intelligent—or perhaps less prone to use or develop our cognitive abilities" (ibid.).

Nicholas Carr (2014) for example illustrates the loss of basic skills due to automation in his book "Glass Cage" with the work of architects, who no longer learn technical drawing independently, but above all learn to use computer-aided design programs. In the process, he says, they lose some of their ability to use spatial imagination, which is exercised precisely in the drawing of geometric figures. In fact, the impact of digital media exposure on the brains of young children can significantly influence language networks in the brain, shaping the development of language processes such as semantics and grammar (Hutton et al., 2020). A similar phenomenon, known to most, can also be observed with AI-supported spelling and grammar aids, particularly when endowed with predictive text functionality—wherein the program anticipates and suggests the subsequent word during the writing process. We react to this support by formulating texts more thoughtlessly, assuming that the program will ensure its appropriateness, and thus walk more and more on digital crutches without yet having the goal of ever putting them down again. Notably, the predictive text feature establishes a robust anchor or benchmark dictating the trajectory of the text, exerting a discernible influence on thought processes by furnishing the "most likely" next word, potentially diverting from the original intended expression.

Prior to the advent of ChatGPT, research and writing procedures constituted a valuable means of cultivating cognitive skills. As we have shown this basic ability to form a judgment, to understand what a judgment is and how we arrive at it requires practice. It is precisely this ability to reason that we should not hand over completely and without consideration, but rather focus on the application of critical thinking and self-reflection as an expression of authorship. Writing and arguing is one of the appropriate means of practicing this skill and applying it in a wide variety of areas. Consequently, this process played a pivotal role in fostering this skill set and, thus, making well-informed decisions. Engaging in a topic via self-generated text pushes us into an "active, persistent, and careful consideration of a belief or supposed form of knowledge in the light of the grounds which support it and the further conclusions to which it tends" (Dewey, 1909). However, the emergence of LLMs may significantly streamline the writing endeavors, raising concerns about the potential impact on the development of reasoning skills. Should students overly depend on information supplied by ChatGPT or analogous tools, there exists a risk of foregoing essential engagement in critical thinking and problem-solving processes vital for cognitive enhancement (Putra et al., 2023). This escalating concern pertains to the potential erosion of peoples' capacity to generate original ideas and construct cogent arguments substantiating their perspectives (Arif et al., 2023).

This pragmatic argument is reminiscent of the *substitution argument*, a concern that can be traced back to the introduction of writing, an apprehension that can already be found in Plato:

"I cannot help feeling, Phaedros, that writing is unfortunately like painting; for the creations of the painter have the attitude of life, and yet if you ask them a question they preserve a solemn silence. And the same may be said of speeches. You would imagine that they had intelligence, but if you want to know anything and put a question to one of them, the speaker always gives one unvarying answer. And when they have been once written down they are tumbled about anywhere among those who may or may not understand them, and know not to whom they should reply, to whom not: and, if they are maltreated or abused, they have no parent to protect them; and they cannot protect or defend themselves."

Plato's point is that the written word cannot respond intelligently, sensitively to the reader. It can be read at will. Ideas can spread quickly, even among those who misunderstand it. The danger may be averted by the interactivity of Chatbots; after all, Chatbots can react and be asked to give explanations once more. Inaccessible passages can be paraphrased, interpretations can be requested. Reasoning can remain in dialogue. The discourse does not break off here. This may be taken as an advantage given the possibility that people may have access to some sort of AI expert.

As the EU points out, these interactive LLMs chatbots can be used in many ways for democratic education due to their engaging design. Chatbots may respond to citizens' questions on candidates' electoral programmes or update citizens on how policies in which they have an interest are evolving. On one hand they can foster civic dialog and on the other hand summarize citizen views and stances to the politicians (Adam & Hocquard, 2023). Interestingly, these suggestions point to the fact that LLMs are to be deployed as deliberative tools rather than as mere decision-making tools (Aichholzer & Rose, 2020). Deliberative tools aim at supporting the understanding of political arguments and gaining insight into political work thereby encouraging citizens to use their power of judgment.

It is pertinent to underscore that skills can undeniably be enhanced through the utilization of LLMs. Particularly, once a skill set has been acquired, it might be deemed imprudent to engage in laborious procedural steps to apply it personally when it could be effectively delegated to a machine just like it seems to be imprudent or at least impractical to carry out complex arithmetic calculations by hand. The pivotal consideration revolves around the timing of task delegation—whether it occurs before or after the acquisition or mastery of a skill by the human agent. Skill loss can of course only occur when a skill was present in the first place. According to Arthur et al. (1998) cognitive skills are especially susceptible to skill loss. Premature and constant delegation may pose potential risks to our societies rooted in humanistic democracy as we explicated. The crux of the matter lies in finding a balance between skill acquisition and the delegation of tasks to machines to attain efficiency. Alternatively, we can frame the question as follows: how can we cultivate a coveted talent, and at what point does it transform into a consistently effective skill, that is, attains a state of being *sufficiently proficient*? Especially the second question is fundamentally an empirical one, necessitating comprehensive research in the years to come. It is, however, important for us to take a proactive stance in navigating this transformative landscape, remaining vigilant of both the risks and benefits that accompany this paradigm shift and ensuring that the foundations of a free and democratic society are left intact.

Authenticity: Whose capabilities and skills are used?

Another central argument to our claim, which we would like to call the *authenticity argument*, can well be illustrated by Heidegger's metaphor of language as the "house of being" (Heidegger, 1971). Language as the "house of being" in which man dwells very aptly describes the significance that language assumes for man in its function of creating meaning and understanding connections. In the uncovering of reason man creates individuality by recognizing herself. We will refer to this relationship and positioning as authenticity: We ascribe authenticity to a person or his or her speech and actions, if their "statements seem genuine, credible and truthful to us, i.e. and true, i.e., not invented, fictitious, just made up or fake." (Luckner, 2007, p.9 - own translation). This refers to the core of authorship and how individuals ascribe attitudes and actions to themselves by always already anticipating that these originate within the individual. Even more, through those deeds and speeches their lives will become unique. This renders discourse intelligible to onlookers as well: acknowledging the autonomy of others in their speech and actions, recognizing that their words and behavior emanate from their own selves. We experience ourselves and the other as being autonomous, as being responsible for our utterances and actions (Nida-Rümelin, 2018; Nida-Rümelin & Weidenfeld 2022). For example, in laborious phases, Nietzsche points out that he even forbade himself to read books, because he did not want anyone's thoughts and voice near him - "it would never occur to me to allow anyone to speak or even to think in my presence. For that is what reading would mean" (Nietzsche, 2009/1908). Nietzsche was afraid that a foreign thought would creep over the mental wall he had erected to protect his genuity - his *sui generis*.

The infusion of Large Language Models into democratic processes may thus raise profound concerns regarding the authenticity that underpins democratic ideals. Authenticity, characterized by statements perceived as genuine, credible, and truthful, is fundamentally challenged by the inherent limitations of LLMs. In democracy, where transparency, trust, and well-informed decision-making are foundational, the potential biases, susceptibility to manipulation, and lack of accountability in machine-generated content pose significant threats. The absence of a human touch, critical thinking, and the inability to trace information back to its sources erode the core tenets of authenticity, thereby challenging the integrity and effectiveness of democratic processes. Striking a balance between the efficiency of automation and the preservation of authenticity is crucial to safeguard the essence of democracy in the face of technological advancements.

As we outlined, the intriguing aspect of large language models such as ChatGPT lies in their remarkable ability to emulate human argumentation. Their generated output exhibits a human quality, rendering their explanations highly persuasive. ChatGPT maintains a conspicuous silence in this regard, as the generated text is intricately intertwined with the specific context, making it deceptively easy for us to misconstrue its outputs as our own work. We appropriate text modules and excerpts from search results, and the responses from ChatGPT give the, albeit wrong, impression that we authored them ourselves. After all, it was our formulation of the pertinent questions that causally led to the very existence of the text.

In other words: the ability of LLMs like ChatGPT, to emulate human argumentation introduces a nuanced challenge to authenticity in democratic discourse. The seamless integration of machine-generated content where only the prompts are human-initiated blurs the lines between genuine authorship and automation. Commonly, we find that the resulting persuasive quality of the generated text, coupled with the intricate contextualization, creates a scenario where users may inadvertently appropriate the content as their own. This not only raises concerns about the authenticity of individual expressions but also challenges the genuine approach to self-positioning and self-discovery in democratic dialogue. The temptation to use generative AI as a shortcut to self-representation introduces a layer of complexity, highlighting the need for careful consideration of the ethical implications surrounding the integration of such technologies into democratic processes.

This phenomenon in many ways contradicts an authentic approach to life: there is no need for us to undergo the process of self-positioning or self-discovery; nevertheless, we can portray ourselves as genuine. Generative AI appears to offer a shortcut to self-representation. This is an enhancement argument, insofar as our human existence is improved by the use of technology. But any enhancement of our capabilities through technology leads, as Coeckelbergh (2013) points out, to a new form of vulnerability. Here, then, we need to think about how our human existence is transformed in its vulnerability in order to evaluate the use of this technology. Vulnerability must be addressed, with a central focus on the philosophical inquiry into the essence of subjectivity and its manifestation.

This goes hand in hand with the question of how we value machine output. Machine-generated output seems to differ from human work due to its inherent lack of authenticity especially when fully disclosed. A notable example is the incident involving Vanderbilt University, where an email of condolence regarding a mass shooting at another university was composed using ChatGPT. The response from people conveyed a sense of betrayal and inappropriateness, highlighting a perceived deficiency in "personal connection and empathy" (Korn, 2023).

Merely formulating a prompt seems insufficient to assert originality when the output relies on text passages crafted by others or machines. We as individuals must try to find arguments and counterarguments and weigh them. We must know where these arguments and the factual knowledge come from, the origins and contexts of statements. We should be aware of when and by whom these statements were made. All this is engulfed within machine-generated output, as if it were unimportant and inconsequential. The knowledge about the genesis of attitudes, however, creates the possibility to be able to position oneself, to come to oneself and to individualize oneself.

Being able to articulate and defend your perspective is a democratic virtue. It involves supporting your views with good reasons, engaging in discussions, and exchanging ideas openly. This necessitates respect and confidence in your own judgment, ultimately contributing to what we would consider an authentic person. This raises a normative query regarding how we intend to manage interactions between humans and machines in this context: To what degree does the output of a machine-human collaboration alter the concept of originality?

In the future, entrusting machines with poetry, text interpretation, literature, journalism, and scientific writing poses a risk. It could lead to a society of mere imitators, confined within a self-referential world. This concern intensifies with the projection that by 2026, 90% of online content will be synthetically generated (Adam & Hocquard, 2023). Moreover, if machines increasingly shape their input based on machine-generated output, humans might relinquish their role in cultural mediation. This involves the compilation, transmission, storage, and dissemination of cultural elements. In our house we find ourselves silenced, to the extent that we might eventually lose the desire to speak at all.

What implications does this have for our comprehension of cooperation predicated on individual autonomy? How can the legitimacy of our democratic society be upheld when its public discourse is no longer driven by individuals but rather by seemingly impartial computer-generated content? For instance, LLMs can be employed to craft responses to regulatory processes or compose letters to the editors of local newspapers, which may subsequently be published (Adam & Hocquard, 2023). Furthermore, the issue extends beyond the matter of copyright pertaining to the training data used, encompassing our approach to handling arguments stemming from human-machine interaction. This necessitates a nuanced proficiency in managing outcomes, their origins, and their evolution, thereby raising the question of how we can cultivate our discernment when engaging interactively with a counterpart whose output can never genuinely express intent. But that is not yet sufficient in the sense of manipulation: Skilful adversarial machine learning, such as data poisoning, constitutes the specific aptitude to manipulate datasets to achieve desired outcomes or deliberately distort results, as the example of Nightshade impressively shows for pictures (Heikkilä, 2023). In regard to democracy this may lead into voter manipulation beyond the fact that we distort the electoral system or election campaigns. Even if the phenomenon is already known from social media (Deb et al., 2019; Aral & Eckles, 2019), the personalized interactive chat function allows for a much more targeted way of responding to users' authorship.

Lost in post factual times? - Teach concepts, not things

What also constitutes part of authenticity and originality points to the aspect that we often think of LLMs as knowledge generators. Nevertheless, we have to treat this argument separately: Let us call this the *facts argument*. Generative AI systems map the state of opinion as data input, which is why they conserve a variation of the data set, but do not extend it. Extensions happen as associations, which are not intended as such, but happen by chance. If these ill-considered outputs extend the state of knowledge then this leads into an unsafe state of facts. However, these skewed and unintended post-factual outputs cannot be readily identified as such.

ChatGPT gives many people the impression that it represents current world knowledge, like comprehensive encyclopedias did in earlier decades and centuries. These attached great importance to source citations and precise evidence, as is still the case today with Wikipedia, while the products of text generation systems do not go through an editorial review process and do not (yet) cite sources (correctly) but instead oftentimes invent fictitious sources - a process widely known as "hallucination" (Rawte et al., 2023). Even if not all knowledge is fully reproduced by people, we can turn to experts and also hold them responsible for wrong statements. In the realm of AI, there seems to be, at least *prima facie*, no identifiable entity to hold accountable for defamation and untrue assertions, neither in a legal nor moral sense. The resolution of this ethical concern requires political and legal decisions. After all, we are familiar with fake news and conspiracy theories from social networks. In this scenario, a solely technical resolution does not seem feasible. In the pursuit of truth, expecting a flawless track record from LLMs is akin to demanding perfection from established institutions like encyclopedias or even the scientific literature. We accept a margin of error in scientific discoveries and, in fact, even strive for falsification since we readily accept that our current knowledge might not prove to be true in the future. We also recognize that encyclopedias, while comprehensive, may undergo revisions and updates or that some facts are in a reasonable dispute (e.g. historical facts). Similarly, understanding the nuanced nature of LLM-generated content and acknowledging the continuous learning process involved allows us to appreciate the information they offer without imposing an impossible standard of infallibility. *Nil satis nisi optimum* is not a standard that we should expect LLMs to achieve, yet, understanding that a LLM is not necessarily a truth generating machine must be seen as a crucial aspect in our interactions with these machines. Frequently, we rely on the quality of a text as a signal of quality. It signals that someone invested effort in creating it, increasing the probability of the text being worthwhile. However, this signal is diminished with LLMs.

Thus, as with any source of information, a discerning approach that considers multiple perspectives and verification methods remains crucial in evaluating a satisfactory threshold for truth. Critically to discuss is the fact that we must be careful not to unlearn to be able to check the outputs of the AI, that we remain able to develop independent thinking and thus are able to check the AI generated texts for plausibility and sense.

Even efforts to label outputs as machine generated do not necessarily alleviate this aspect. Ultimately, these markers can themselves be manipulated, and the handling of the output may still not change, allowing the misinformation to spread. Once it is in circulation, containment becomes a challenging endeavor. Politicians often interpret correspondence received from their constituents as a reflection of public sentiment that can inform their decision-making - even if it is a new form of publicity. Furthermore, with the emergence of artificial intelligence, a concerning development known as astroturf campaigning has become notably easier. This strategy makes it possible for a small, disguised group posing as a legitimate grassroots movement, disseminating a biased representation of public opinion. Additionally, AI has the potential to foster the illusion of widespread political consensus by flooding online platforms with millions of automatically generated posts on a given subject (Adam & Hocquard, 2023). As LLMs become increasingly personalized and interactive, there is a heightened risk that people may not critically assess the information they encounter or fully understand its origins.

## Education as a cornerstone of democracy

As previously emphasized, the core of the democratic lifestyle is found in the interchange among individuals articulating their reasoning and collectively reaching agreements. This exchange, grounded in mutual respect among free and equal individuals, extends across all aspects of our shared existence. This does not imply a uniformity of interests, preferences, or desires among us, nor an obligation to adopt them as our own. Rather, it suggests expressing them in a manner that upholds and safeguards the autonomy and equality of those who hold contrasting views. This, in turn, is based on our ability to exercise our own - authentic - power of judgment. Practice aligns with skill. The power of judgment needs to be trained to apply rules of understanding and principles of reason on singular events. Introducing Plato once more:

"Is not rhetoric, taken generally, a universal art of enchanting the mind by arguments; which is practiced not only in courts and public assemblies, but in private houses also, having to do with all matters, great as well as small, good and bad alike, and is in all equally right, and equally to be esteemed-that is what you have heard?"

Hence, as we aim to convey, active participation in argumentation appears crucial for inclusion and engagement in various social spheres. Articulating, expressing, and standing for one's perspective is simultaneously a vital aspect of personal authenticity. Furthermore, it is a skill imperative for all individuals to navigate and thrive in a democratic society, promoting equality and freedom.

In a world in which AI-driven machines increasingly impact our everyday lives, the question thus arises as to where we should create spaces for independent expressions of thought and action - in which we should develop, teach and learn the activity of reasoning. Leveraging Large Language Models undoubtedly holds promise for enhancing educational outcomes, as posited by Kasneci et al. (2023), who delineate advantages encompassing heightened student engagement, interactive learning, and personalized educational experiences, among various potential enhancements. A study by Baidoo-Anu & Ansah (2023) provides suggestions on optimizing the utilization of ChatGPT for the enhancement of teaching and learning outcomes. Nevertheless, this study addresses the tangible danger associated with an excessive dependence on artificial intelligence, coupled with the inherent human inclination to opt for the path of least resistance. Krügel et al. (2023) find that humans who are supposed to correct bad machine decisions oftentimes do not and instead turn into "partners in crime" (Krügel et al. 2023). In the event that LLMs are not judiciously employed as a constructive augmentation but are excessively depended upon for outsourcing cognitive processes, a critical question arises: How can we guarantee that individuals acquire the requisite skill set essential for becoming capable and autonomous citizens within a democratic state? Again, this issue pertains to the temporal aspect of task delegation, specifically, whether it transpires prior to or subsequent to the human agent's acquisition or mastery of a skill.

This question then has particular implications for our education system: How are we supposed to train children and young people to reason if they no longer have to argue independently, i.e. without the help of a machine? When opting for the easiest route involves delegating reasoning, the consequence is a potential decline or lack of development in our own capabilities. Cognitive skills such as imagination, creativity, and critical reasoning could suffer from excessive use of AI, as we are no longer forced to put our thoughts down independently. Formulating thoughts in an orderly fashion, however, is the vehicle of the power of judgment, as it forces us to organize ideas and concepts and cast them into a form. American writer Flannery O'Connor summed up this human experience when she describes: "I write because I don't know what I think until I read what I say" (O'Connor & Fitzgerald, 1979), The more text work we delegate to AI, the less powerful the connection and realization of our thinking through writing. In our function as editors or controllers we need concepts, terms and rules according to which we can categorize what we perceive. Drawing good conclusions for our actions from applying explicit and implicit rules and facts in accordance.  At the same time, this points to the fact that language manifests itself in syntax, semantics, and pragmatics, and that normative requirements are grounded in the life-world.

ChatGPT is often hailed as a groundbreaking tool for human-like responses and reactions. But ChatGPT does not have semantics, does not understand meaning and contexts, even if suggested by a language model based on the probability of word sequences. This concept is similar to the situation in the thought experiment of the so-called "Chinese Room" by John Searle. In Searle's scenario, he envisions being in a room with a computer program guiding him on responding to Chinese characters slipped under the door. Despite lacking any understanding of Chinese, he mechanically follows the program, manipulating symbols and numerals like a computer. As a result, he sends out fitting strings of Chinese characters under the door, creating the illusion for those outside that a Chinese speaker occupies the room. According to Searle (1980), the thought experiment highlights that computers simply apply syntactic rules to handle symbol strings, lacking genuine comprehension of meaning or semantics. Similar to someone learning a foreign language based only on strings, ChatGPT can process strings without an understanding of their meaning. In the context of Searle's Chinese Room thought experiment, the difference of LLMs lies in the learning approach. While the model doesn't follow explicit syntactic rules, it grasps intricate language structures through the patterns it discerns during training. It's a nuanced distinction, as the model learns not by being explicitly programmed with rules but by implicitly capturing the underlying patterns within the data it processes. Thus, ChatGPT can play chess, but does not understand the rules, so cannot always apply rules of the game to individual cases. ChatGPT captures its data tokens and because of its stochastic model makes moves with non-existent pieces. These moves are understandable in themselves, but are wrong in terms of the game, in terms of an entire set of rules. So to speak it cannot configure the single case since it cannot transcend its rules-based determination.

Readers must therefore interpret the strings themselves and establish a context of meaning. They understand the meaning of the output only because they have learned the usage of words within contexts and the sequences of actions through their participation in daily routines: They can grasp semantics and pragmatics and corroborate them through their own reason, thoughts, judgments and actions. People can anticipate and comprehend why others follow rules. They can think, speak, and act together because they know what the thought, speech act, or action is intended to accomplish. In this way, by anticipation we attribute meaning and reason to our fellow human beings.

When it comes to educating children, youth, and even adults, it's crucial to elucidate these aspects of generative AI. We need to convey the message that machine-generated output doesn't equate to reasoning, comprehension, or judgmental prowess (Larson, 2023). We can no longer trust AI machines as easily as we trust a calculator. In any case, the output requires a careful check on our part. Not like calculators, whose output we rarely (have to) question. This requires sharper judgment when it comes to digital. The debate here is about digital literacy, which new or expanded skills we should have when dealing with digital technology.  In this context, Shannon Vallor (2016) speaks of the techno-moral virtues, to which she adds techno-moral wisdom to the classic virtues such as modesty or courtesy. What is initially pleasing is that many of our desirable attitudes and lifestyles remain intact even when dealing with digital. She formulates a techno-moral wisdom that combines a multitude of virtues in order to achieve the good: "Deliberative skill alone is not practical wisdom, since the latter integrates the full range of moral habits [...]. Skill in choosing the best means to a perceived good is meaningless unless I am habituated to aim at the correct good in the first place, and it isn't virtuous unless I am motivated to choose that good for its own sake. A person who skillfully selects means to attain ends that are not actually good ends for a human being to seek, or who aims at the right ends but for the wrong reasons, is merely clever, not wise." When dealing with ChatGPT, this first means that users have to understand how these systems work. You must have a little technical literacy (Hofmann et al. 2019). You need to understand how input relates to output. This is the only way we can learn to understand how we can handle the output responsibly, i.e. wisely. People are frequently unaware that their behavior serves as input for AI systems.

ChatGPT cannot be reasonable. However, ChatGPT can be developed, deployed and used responsibly. The objective should be to improve human capabilities regarding practical wisdom rather than replacing it, always keeping in mind that our democratic values are necessary to pursue. Yet, transparency about the AI does not seem to be enough. Recent studies have shown that people, regardless of being informed about whether advice originated from a machine or a human, exhibited unchanged behavior (Krügel et al., 2023; Krügel et al. 2023a) and can sometimes "hide behind" machine decisions (Feier et al., 2022). This highlights the significance of promoting digital literacy rather than solely emphasizing transparency. However, the responsibility does not rest solely on the user; developers of AI must also weigh ethical considerations in the development of their products (Gogoll et al., 2021). Hence, incorporating ethical considerations into the AI development process is essential, ensuring the contemplation and implementation of good design right from the outset (Zuber et al., 2022).

The facts of machine limitations show us the skills that we must constantly learn. Thus, we can use ChatGPT's output, which becomes the content of our experiential world, to exercise our mind and judgment. We are constantly instructed by ChatGPT to self-critically question what we perceive, which is why machines like ChatGPT can perhaps be used as machine-assisted Socratic dialogues. Ultimately, the foundational humanistic and Enlightenment ideals of independent reasoning, thoughtful contemplation, and purposeful output necessitate a guided practice in discourse. Under the guidance of a teacher, children and adolescents can engage with LLMs or generative AI on a topic until their reasoning skills in that area are improved through practice. Eventually, they can derive more creative conclusions than their computer counterparts - and notice this with self-confidence.

Digital machines cannot check the criteria of successful communication themselves and are blind to more subtle, often ironic forms of rule transgression. Connotations that enable people to let things slide sometimes, what can be called ambivalence tolerance, are evaluated as errors in a calculation. In our everyday lives, however, this is precisely what distinguishes us: We praise people for their tact. This is the skill we teach children, whether we have calculators or operate speech machines. It is precisely for this practiced thinking that we still need the other, semantic knowledge as normative reference points, because in the absence of a railing, there is no stable footing, and without critical examination, coherent practice is unattainable.

Learning and practicing judgment is a slow and arduous task. It is not absorbed in factual knowledge, nor in the explicit presentation of rules. In democratic societies, however, it is an essential component: thinking, absorbing, weighing and judging together cannot be delegated to AI. Individual processes of democratic life may be. As Margo Loor (2023) wisely puts it:

"Voting is quick. This can be done on a screen. This is not democracy; done wrong it can be the tyranny of the majority. Voting may be part of the democratic process, but it does not replace the core of democracy, which is human thinking. First of all, everyone must think about society's problems and then think about and discuss the issues with others. This must be followed by listening to the arguments of the various stakeholders, both majorities and minorities, and only then can you vote."

30Putra, F. W., Rangka, I. B., Aminah, S., & Aditama, M. H. (2023). ChatGPT in the higher education environment: perspectives from the theory of high order thinking skills. *Journal of Public Health*, fdad120.

Rawls, J. (1993). Political liberalism. New York: Columbia University Press.

Rawte, V., Sheth, A., & Das, A. (2023). A survey of hallucination in large foundation models. *arXiv preprint arXiv:2309.05922*.

Rowlands, I., Nicholas, D., Williams, P., Huntington, P., Fieldhouse, M., Gunter, B., ... & Tenopir, C. (2008, July). The Google generation: the information behaviour of the researcher of the future. In *Aslib proceedings* (Vol. 60, No. 4, pp. 290-310). Emerald Group Publishing Limited.

Searle, J. R. (1980). Minds, brains, and programs. Behavioral and Brain Sciences3: 41724.[aSL](1992) The rediscovery of mind.

Stahl, B. C., & Eke, D. (2024). The ethics of ChatGPT–Exploring the ethical issues of an emerging technology. *International Journal of Information Management*, *74*, 102700.

Strawson, P. F. (1974) [2008]. *Freedom and resentment and other essays*. Routledge.

Vallor, S. (2016).*Technology and the virtues: A philosophical guide to a future worth wanting*. Oxford University Press.

Wedgwood, R. (2017). *The Value of Rationality*. Oxford, New York: Oxford University Press.

Wittgenstein, Ludwig (1953). Philosophical Investigations: The German Text, with a Revised English Translation. Blackwell. pp. § 23. ISBN 9780631231592.